# Glare suppression by coherence gated negation

Edward Haojiang Zhou[*,1], Atsushi Shibukawa[1], Joshua Brake[1], Haowen Ruan[1], Changhuei Yang[1]

[1]*Electrical Engineering, California Institute of Technology, 1200 East California Boulevard, Pasadena, California, 91125, USA*

*Correspondence to Edward Haojiang Zhou at hzzhou@caltech.edu.

Imaging of a weak target hidden behind a scattering medium can be significantly confounded by glare. We report a method, termed coherence gated negation (CGN), that uses destructive optical interference to suppress glare and allow improved imaging of a weak target. As a demonstration, we show that by permuting through a set range of amplitude and phase values for a reference beam interfering with the optical field from the glare and target reflection, we can suppress glare by an order of magnitude, even when the optical wavefront is highly disordered. This strategy significantly departs from conventional coherence gating methods in that CGN actively 'gates out' the unwanted optical contributions while conventional methods 'gate in' the target optical signal. We further show that the CGN method can outperform conventional coherence gating image quality in certain scenarios by more effectively rejecting unwanted optical contributions.

The ability to optically illuminate and image a target hidden behind a scattering medium is important in many applications, including transportation, remote sensing, biomedicine and astronomy. A classic example is the scenario of driving through fog at night with the automobile headlights on. The degradation of image quality in such scenarios can be generally ascribed to two effects: the optical wavefront distortion caused by the scattering medium and the glare from light backscattered from the scattering medium. The wavefront distortion limits our ability to perform diffraction-limited imaging and optical focusing. However, even in cases where the wavefront distortion does not prohibit imaging of the target, the sheer intensity of the glare can mask the weak optical reflection from a target and thereby prevent us from observing the target altogether.

Recent developments in wavefront shaping and adaptive optics have shown great promise in addressing the wavefront distortion challenge[1–6]. These methods have improved the imaging resolution beyond what was thought possible even a decade ago. However, in almost all of the demonstrations performed so far, the problem of glare is averted either by choosing a target that emits light at a different wavelength (fluorescence[7] or second harmonic generation[8]) or by designing the experiments to operate in a transmission geometry. Glare remains a challenge largely unaddressed in the context of these developments. Unfortunately, glare is unavoidable in a variety of practical scenarios - driving in a foggy night is a good example. In that scenario, the objects you would like to observe are unlikely to be fluorescent, and you simply cannot rely on having an independent light source behind the objects to provide you with a transmission imaging geometry.

Glare suppression in principle is possible using of time-of-flight methods with the help of fast imaging systems, such as those based on intensified charge-coupled device (ICCD) technology[9–11] or single photon avalanche diode (SPAD) arrays[12–14]. These devices are able to bin the light arriving at the detector with fine temporal resolution and therefore glare can be suppressed by discarding glare photons selected by their arrival time. Unfortunately, these instruments are very costly. But perhaps more importantly, the range to which they can suppress glare is determined by their response speed. The best commercial instruments available have a response time of 0.5 ns, which translates to a minimum length of ~10 cm for which they can suppress glare by time gating. Recently, SPAD arrays of temporal resolution of 67 ps have been demonstrated. However, they are currently only available in small array sizes (32 x 32 pixels)[13,15].

There have also been some interesting developments on the use of modulated illumination and post-detection processing in the phase or frequency domain to achieve time-of-flight based gating[16,17] One significant limitation to these methods is that they need to contend with glare associated noise, as the glare is not suppressed prior to detection. Moreover, such techniques are limited by the frequency bandwidth of the sensors, which leads to a minimum length involved on the order of meters. This length limitation for all known glare countering methods precludes useful applications of such time-of-flight methods in biomedicine where the length scale of interest ranges from microns to millimeters.

The streak camera is yet another fast response optical detection system. Its response speed is on the order of one picosecond. Unfortunately, the streak camera is intrinsically a one-dimensional imaging system. Recently, it has been demonstrated that the use of compressed sensing can allow the streak camera to perform fast two-dimensional imaging[18,19]. However, the object sparsity constraint is too restrictive for the majority of glare suppression applications.

Here, we report a method, termed coherence gated negation (CGN), that is capable of coherently suppressing glare through the use of destructive interference to allow improved imaging of a weak target. This method can operate over a length scale span that is limited only by the coherence length of available optical sources, which can range from microns (for superluminescent diodes) to kilometers (for fiber lasers). CGN shares its roots with acoustic noise cancellation[20]. The basic idea is to use a reference optical field of the same magnitude and opposite phase to destructively interfere with the glare component of a returning optical field to null out the glare and its associated noise, thereby allowing the electronic detector to measure only the optical signal from the hidden target. In the case of acoustic noise cancellation, the amplitude and phase of the unwanted signal can be separately measured and used as input in the cancellation process. In CGN, we do not have this luxury as we do not have prior knowledge of the glare optical field characteristics. Here, we instead employ a light source of suitable coherence length such that a) the glare optical field is coherent with the reference optical field, and b) the target reflection is incoherent. By permuting through a specific set of amplitude and phase values for the reference field, we ensure that the condition for effective destructive interference is met within a certain error bound for one of the permutations. By screening for the minimum detected optical signal through the whole set, we can then determine the signal reflected from the target. When performed in an imaging context, this allows us to use a single permutation set operating over all the camera pixels at once to generate a glare suppressed image even if the optical field is highly disordered and speckled.

Using this approach, we experimentally demonstrate the ability to suppress the glare by a factor of 10 times with the use of a permutation set of size 256. Our experimental design choice also allowed us to demonstrate glare suppression on the length scale of 2 mm - a regime that conventional time-of-flight methods are presently unable to reach. Finally, we discuss the advantages and tradeoffs of CGN versus traditional coherence gating methods and report our experiments demonstrating CGN's ability to image targets at different depths without system alterations, and several scenarios where CGN can provide better target image quality than conventional coherence gating methods.

## Results

**Principles**

A concise setup to explain the principle of CGN is shown diagrammatically in Fig.1. A laser beam illuminates a two-dimensional target located behind a scattering sample. The returning light, which consists of light that is back-scattered by the scattering medium as well as light reflected from the target, is captured by the imaging system, resulting in an image of the target obscured by glare. On the camera sensor chip, the captured optical field is the superposition of the glare $\mathbf{E}_{glare}(p,q)$ and the target reflection $\mathbf{E}_{target}(p,q)$, where $p$ and $q$ are the pixel numbers in the $x$ and $y$ direction, respectively. To realize CGN, a collimated reference beam $\mathbf{E}_{ri}(p,q)$ is added on the camera by a beamsplitter to interfere with $\mathbf{E}_{target}(p,q)$ and $\mathbf{E}_{glare}(p,q)$. We perform path length matching of the glare contribution and the reference beam. By choosing the coherence length of the laser source appropriately, we can make sure the glare contributions from the extended scattering medium are in coherence with the reference beam. As long as the optical path length of the target reflection is substantially different from the majority of the optical path lengths of the glare components, the target reflection will not be in coherence with the reference beam. We then permute the reference beam through a series of phase and amplitude changes. The observed image intensity for the $i$th image $\mathbf{I}_i(p,q)$ can be expressed as

$$\mathbf{I}_i(p,q) = \mathbf{I}_{target}(p,q) + |\mathbf{E}_{glare}(p,q) + \mathbf{E}_{ri}(p,q)|^2, \tag{1}$$

where $\mathbf{I}_{target}(p,q) = |\mathbf{E}_{target}(p,q)|^2$ is the target intensity.

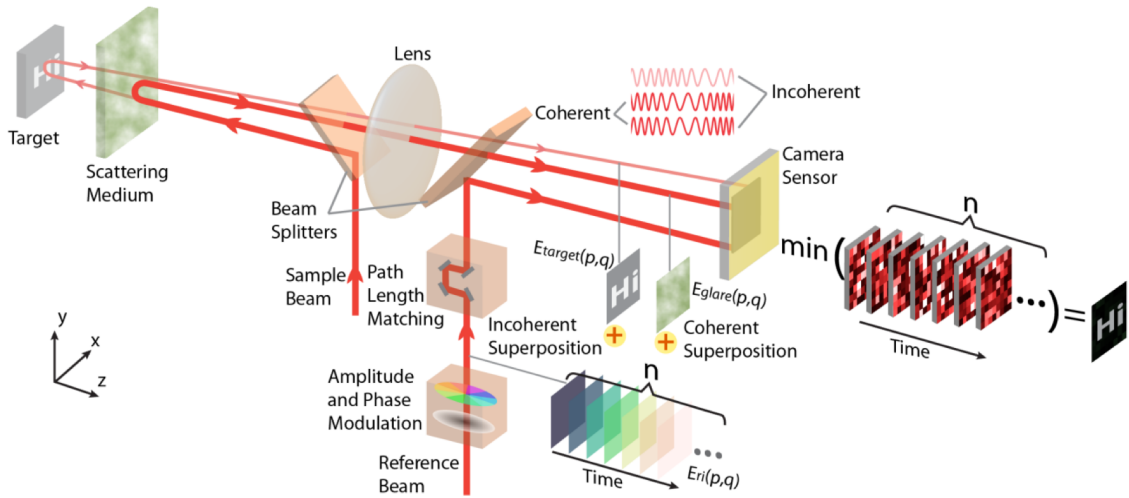

**Figure 1 | Principle of CGN technique.** The CGN system uses a laser as the illumination source for the active imaging system. With the presence of a scattering medium, a significant portion of the light is backscattered to the camera that images the target. A plane-wave reference beam, with path length and polarization matched to the backscattered light (glare), is used to cancel the glare by destructive interference. In this case, we step both the amplitude and phase of the reference beam to cover a significant dynamic range of the glare and combine each of them with the glare respectively, resulting in a set of speckle images from the camera. By taking the minimum intensity of each pixel vector along the time axis of the speckle image set, we can reconstruct the image of the target with significant glare suppression.

We further assume that the imaging is performed in such a way that the image speckle size is greater than the camera pixel size. This ensures that there are no phase variations across the surface of any given pixel. In this case, the minimum value that $\mathbf{I}_i(p,q)$ can take is $\mathbf{I}_{target}(p,q)$, which occurs when $\mathbf{E}_{ri}(p,q)$ is of the same magnitude and opposite phase of $\mathbf{E}_{glare}(p,q)$ (destructive interference), that is $|\mathbf{E}_{glare}(p,q) + \mathbf{E}_{ri}(p,q)|^2 = 0$. As such, by permuting through different phase and amplitude values for $\mathbf{E}_{ri}(p,q)$, we can determine $\mathbf{I}_{target}(p,q)$ for each image pixel simply by taking the smallest measured $\mathbf{I}_i(p,q)$ through a set of reference field permuted images. As the glare cancellation is performed in the optical regime, CGN can allow detection of the target without any noise consideration from the glare at all.

In practice, we do not expect complete destructive interference to occur as the glare optical field's phase and amplitude are continuously distributed, while the modulation of the reference phase and amplitude can only be performed in a discrete fashion. The greater the permutation set, the more effectively we can suppress the glare at the price of longer data collection time.

**Experimental demonstration of glare suppression with CGN**

To validate the CGN method, we implemented the experimental setup shown in Fig. 2a. A continuous-wave laser (MGL-FN-532, Opto Engine, 532 nm wavelength, ~1mm coherence length) was used as the light source. Light from the laser was split into a reference and sample beam by a beamsplitter (CBS). The sample beam illuminated the target, which was placed 2 mm behind the scattering sample (SS) (shown in Fig. 2a). The scattering sample (15 mm ($x$) × 25 mm ($y$) × 1 mm ($z$)) consisted of polystyrene particles (3 μm in diameter) in gel phantom (concentration $6.8\times10^7$ ml$^{-1}$, see Methods, sample preparation). The back-reflected light consisted of reflections from the target and glare from the scattering sample. On the other optical path, the reference beam was passed through an amplitude and phase modulator, spatially filtered, and collimated into a plane wave. The collimated reference beam illuminated the camera sensor chip at normal incidence. The reflected light from the target and the glare propagating through BS1 was captured by an objective lens (OBJ), filtered to a single polarization, and imaged by a tube lens (L1) onto the camera. The optical field's effective angular range was 6.3 degrees. This translates to an optical speckle spot size of 19.2 μm at the sensor. In comparison, the camera pixel size is 4.4 μm. This allowed us to enforce the CGN operating requirement that the phase not vary substantially across any given pixel's surface. By path length matching, the collimated reference beam only interfered with the glare but not the reflection from the target. Before CGN was applied, an optical shutter (OS) blocked the reference beam, and an image of the target occluded by glare was captured as shown in Fig. 2c. The optical shutter was then opened and CGN applied. The reference beam was modulated through all permutations of 8 amplitude values and 32 phase values successively. After the reference beam went through all the permutations, a glare suppressed CGN image was acquired (Fig. 2d). Comparing the image before CGN (Fig. 2c) and after CGN (Fig. 2d), we can clearly discern the previously obscured target. To quantify the glare suppression ability of the CGN technique, we define the glare suppression factor as the ratio between the mean intensity of the glare before and after the CGN process. Through a null target experiment, we determined that the glare suppression factor was ~10 for this experiment. Unsurprisingly, the glare wavefront was highly disordered. The glare wavefront as determined by the CGN process is reported in the Supplementary Information.

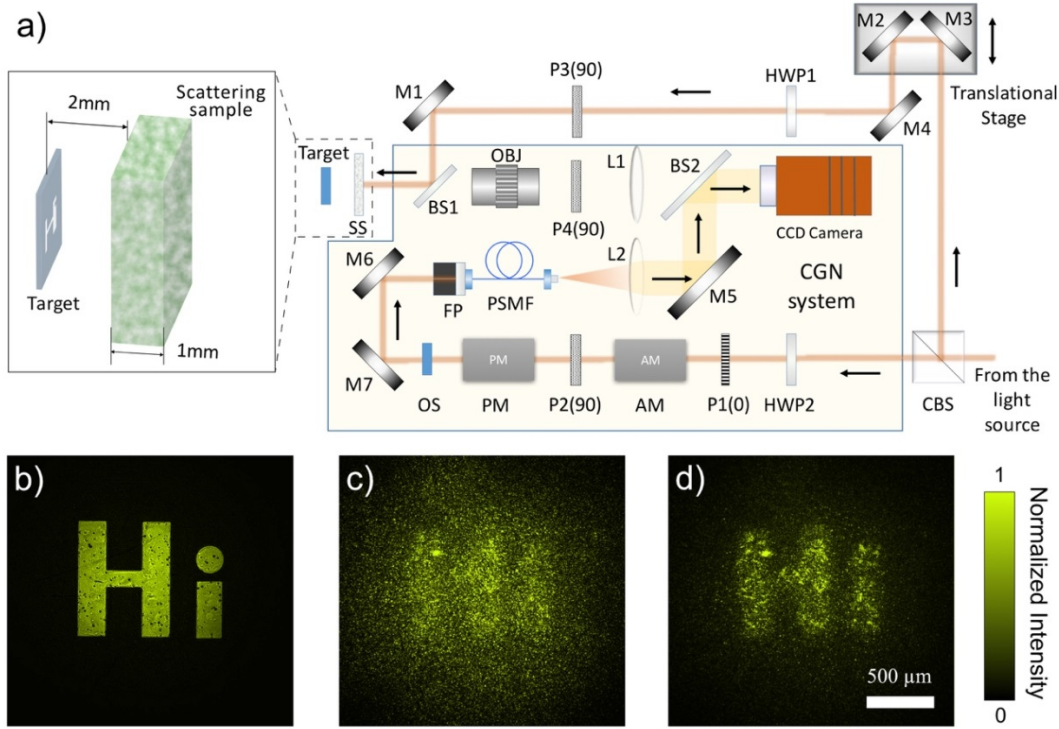

**Figure 2 | Experimental demonstration of CGN.** (a) Experimental setup. (b) Image of the target without glare. (c) Image of the target with glare before CGN. (d) Image of the target after CGN.

As discussed earlier, the glare suppression factor is directly determined by the size of the permuted set of reference amplitude and phase values. We next performed an experiment to measure the glare suppression factor with different numbers of steps in the reference field phase and amplitude. To eliminate the influences of laser coherence for residual glare intensity, a laser with a long coherence length (Excelsior 532, Spectra Physics, 532 nm wavelength, >9 m coherence length) was used in this experiment. A series of glare suppression factors were measured through CGN experiments with a null target but the same scattering medium (15 mm ($x$) × 25 mm ($y$) × 1 mm ($z$)) consisting of polystyrene particles (3 μm in diameter) in a carrageenan gel phantom (concentration $6.8×10^7$ ml$^{-1}$, see Methods, Sample preparation). We varied the number of amplitude steps from 1 to 10 and the number of phase steps from 1 to 32. The full chart is shown in the Supplementary Information. The plots of selected combinations are included in Fig. 3a. For comparison, the expected CGN factor computed through an idealized simulation are shown as well (see Supplementary Information for details). The mismatch between the measured and ideal CGN factor can be attributed to: a) phase jitter in the reference beam and sample beam due to vibration in the system, b) noise in the electronics including the laser and electro-optical modulator, and c) limited extinction ratio of the amplitude modulator and polarized optics, etc. Fig. 3b shows a histogram of the glare intensity before and after CGN for the situation where we permute through 10 amplitude steps and 32 phase steps. In this case, we experimentally achieved a glare suppression factor of ~30.

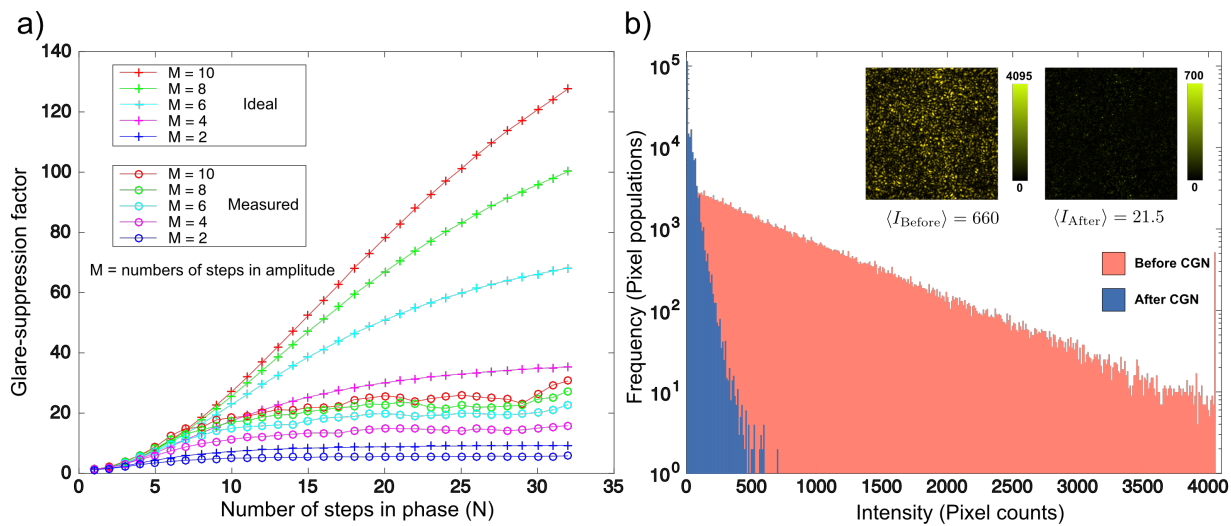

**Figure 3 | Characterization of glare suppression factor.** (a) Comparison of glare suppression factor between measurement and simulation results with various phase and amplitude steps. (b) Histogram of pixel intensities before and after glare suppression, with intensity maps of the glare shown in the insets.

## Comparison to coherence gating

By detecting only the optical field component that is coherent with the reference field, conventional coherence gating methods can also reject glare. However, the way in which conventional coherence gated (CG) and coherence gated negation (CGN) imaging methods work are opposite in nature. While CG imaging methods are good at 'gating in' an optical field originating from a specific chosen distance, CGN is good at 'gating out' the glare optical field. These different approaches to imaging in the presence of scattering and glare lead to two key distinctions between conventional CG methods[21–23] and the CGN approach.

The first key distinction between CG and CGN is that CG methods reject glare contributions as well as any other potential optical signals of interest outside the coherence window. In comparison, CGN can permit detection of all optical signals that do not share the same coherence window as the glare components. This distinction is practically important. In a scenario where there are two objects at different distances behind a fog, a CG method, such as coherent Light Detection And Ranging (LiDAR), is only able to detect one object at a given time. Another class of CG methods, based on spectral sweeping, such as swept source optical coherence tomography[22], can perform simultaneous depth-ranging of multiple objects. However, such methods are intrinsically limited in their range span. Moreover, if the objects' distances are unknown, the coherent LiDAR system would have to be exhaustively range-scanned to find the objects. In comparison, by working to suppress glare, CGN permits direct observation of all objects at any range beyond the glare suppression region. However, this advantage does come with a compensating disadvantage - CGN is not capable of providing depth information of the objects.

To demonstrate CGN's advantage over CG in this aspect, we performed the following experiment. As shown in Fig. 4a, following the aforementioned procedure, CGN was applied to the target located at different positions A, B and C, which correspond to 1 mm, 2 mm, and 3 mm behind the scattering sample, respectively. Since CGN works by coherently gating out the glare component of the light, no adjustment is required to adapt to the depth change of the target, as long as the target remains within the depth of field of the imaging system. The experimental results are displayed in Fig. 4b-g. Fig. 4b-d are images of the target captured before glare suppression, while Fig. 4e-g are images captured after glare suppression. From their comparison, we can easily discern that glare is suppressed and the visibility of the target is enhanced.

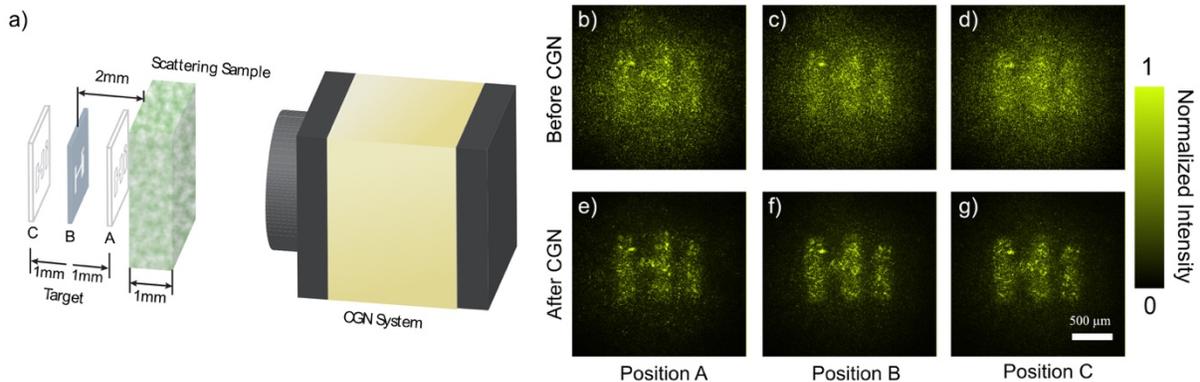

**Figure 4 | Reconstruction of the target at different distances.** (a) Illustration of the target positions. (b-d) Before CGN, images of the target at position A, B and C, respectively. (e-g) After CGN, images of the target at position A, B, C, respectively.

The second key distinction between CG and CGN is that if an element contributing glare and a weak target object both lie within the coherence envelope of the light source, CGN can actually provide a superior signal-to-background image of the object. To clearly and simply explain this point, we consider a scattering sample as the glare contributor and a weak target placed at a distance L away from the CGN system. Here the coherence length of the light source is C, and L is set to be shorter than C. Under CGN operation, we adjust the path length to match the reference beam with the glare contribution. CGN will completely suppress the glare in this situation. As the target is partially coherent, we would expect a diminished signal associated with the target as only the incoherent portion of the target will contribute to image formation. In contrast, under conventional CG operation, we would match the reference beam path length to the target. This results in the detection of the target as well as a partial contribution from the coherent component of the glare. In aggregate, the CGN detection scheme results in a depressed target signal with no glare background, which is more desirable than the CG case where a glare background is present. This result is also valid over the range of an extended scattering media.

To demonstrate CGN's advantage, we performed the following experiment. As shown in Fig. 5a, a thin scattering medium (15 mm ($x$) × 25 mm ($y$) × 0.5 mm ($z$)) consisting of polystyrene particles (3 μm in diameter) in a gel phantom (concentration $6.8×10^7$ ml$^{-1}$, see Methods, Sample preparation) was attached directly on the top of a reflective target. CGN was applied after the path length of the reference beam was matched with the glare as shown in Fig. 5b. Images of the target acquired before and after CGN are included in Fig. 5c and Fig. 5d, respectively. After these images were acquired, the path length of the reference beam was adjusted to match the reflection from the target and phase shifting holography[24] was applied as a demonstration of a CG approach. The retrieved intensity map from this procedure is shown in Fig. 5e.

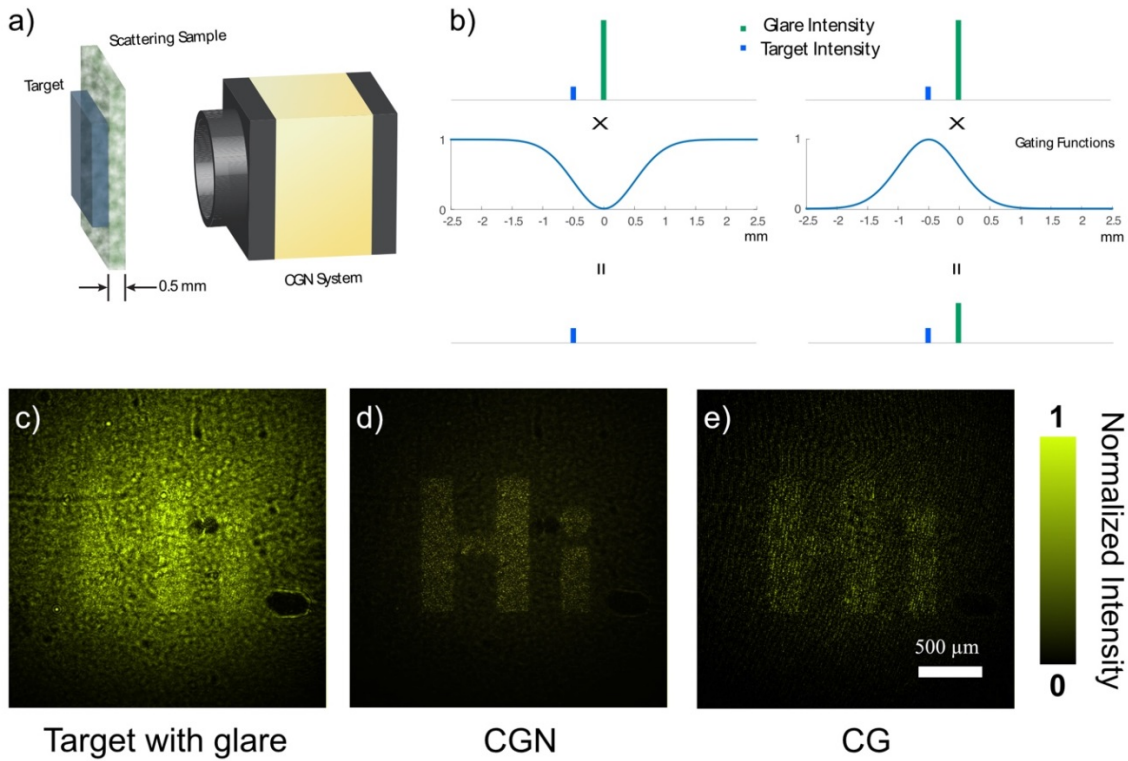

**Figure 5 | Comparison of CGN and CG techniques.** (a) Illustration of the experimental configuration. (b) Diagrams that illustrates the difference between CGN and CG techniques when both the target and scattering

medium are within the coherence gating window. The CGN technique uses an inversed coherence gating function to gate out the glare significantly, with less suppression of the target, resulting in higher target intensity than glare. The CG technique gates in the target intensity with less preservation of glare. However, the residue of the glare remains higher than the target intensity because of the strong nature of the glare. (c) Original image of the target with glare. (d) Reconstructed image of the target with the CGN technique. (e) Reconstructed image of the target with the CG technique.

## Discussion

In this series of experiments we demonstrated the differences and advantages of CGN compared to hardware based time-of-flight glare reduction systems and conventional coherence gating methods. CGN's ability to suppress glare over optical distances as short as several microns through the use of low coherence light sources, such as superluminescent diodes, contrasts favorably compared to conventional time-of-flight hardware. We also showed that, by suppressing glare and permitting all other optical signals to pass, CGN allows for the simultaneous imaging of objects at different distances. In contrast, CG methods are good at imaging objects at a given distance and rejecting optical contributions before and after the chosen plane. We further showed that CGN can outperform CG methods in image quality under certain conditions - specifically, when the glare components and the target optical field are within the same coherence window of the interferometer.

The CGN design for a specific application will be application dependent. For example, in the scenario where we would like to cancel glare from a fast changing scattering medium, we would likely need both a fast camera and a fast reference field permutation apparatus. Alternately, in this scenario, we may instead choose to perform CGN in a pixel-by-pixel basis rather than a full-frame basis. For pixel-by-pixel CGN, we would focus on a single pixel and iteratively derive the correct reference cancellation field quickly. In an ideal situation, we would only need a few measurements to arrive at the correct field[25,26]. By performing CGN this way, we can progressively work through all the image pixels. As long as the time taken to optimize glare suppression for each pixel is shorter than the time scale at which the scattering medium is decorrelating its optical field, we can expect to suppress glare effectively.

## Methods

### Sample preparation

Polystyrene microspheres with a mean diameter of 3 μm (Polybead Microsphere, Polysciences, Inc.) were mixed with a 1.5% carrageenan gel in aqueous phase. The mixture was cast in a mold of size 15 mm × 25 mm, with a thickness of 1 mm or 0.5 mm. The medium had a theoretically calculated scattering coefficient of $\mu_s=\sigma_s N=1.3$ mm$^{-1}$, where the density of the microspheres $N$ is $6.8\times10^7$ml$^{-1}$ and the scattering cross section $\sigma_s$ is 18.7 μm$^2$.

The target was made by attaching a positive mask showing letters "Hi" to an optical mirror. The height of the letter 'H' was 1 mm.

**Setup**

The experiment was carried out on a custom-built setup as depicted in Fig. 2a. A continuous-wave laser (MGL-FN-532, Opto Engine) with 532 nm wavelength and a coherence length of ~1mm was used as the light source to illuminate the target. A laser with long coherence length (Excelsior 532, Spectra Physics, 532 nm wavelength, >9 m coherence length) was used only for characterizing the glare (Fig. 3). Light from the laser was split into a reference beam and a sample beam by a beamsplitter (CBS). The sample beam illuminated the target at 2 mm behind the scattering sample (SS) (shown in Fig. 2a). Light reflected from the target and the glare propagating through a beamsplitter (BS1) was captured by an objective lens (OBJ, M Plan Apo 2×, NA=0.055, Mitutoyo), linearly polarized, and imaged by a tube lens (L1) on to the camera (Resolution: 1936 (H) × 1456 (V), Pixel size: 4.54 μm x 4.54 μm, Prosilica GX, Allied Vision). The optical field's effective angular range was 6.3 degrees. This translates to an optical speckle spot size of 19.2 μm on average at the sensor (pixel size 4.4 µm). The reference beam was modulated by an amplitude modulator (EO-AM-NR-C4, Thorlabs) and a phase modulator (EO-PM-NR-04, Thorlabs) through permutations of 8 amplitude steps and 32 phase steps successively. The polarization direction of the reference beam was aligned with the sample beam. The reference beam was spatially filtered, collimated into a plane wave, and coupled to the camera in normal direction using a beamsplitter (BS2). The path length of the reference beam matched with that of the light reflected from the scattering sample.

## Acknowledgements

The authors would like to thank Dr. Mooseok Jang and Ms. Michelle Cua for helpful discussions. This work is supported by the National Institutes of Health (1U01NS090577-01) and a GIST-Caltech Collaborative Research Proposal (CG2016). J.B. acknowledges support from from the National Institute of Biomedical Imaging and Bioengineering under a Ruth L. Kirschstein National Research Service Award (1F31EB021153-01) and from the Donna and Benjamin M. Rosen Bioengineering Center.

## Author contributions

E.H.Z and C.Y. conceived the idea, with helpful discussions from A.S., J.B., and H.R. E.H.Z., A.S., and C.Y. designed the experiments, which were carried out by E.H.Z. A.S. fabricated and characterized the samples. E.H.Z., A.S., and H.R. analyzed the data, developed the theoretical background, and carried out the simulations. All the authors contributed to the preparation of the manuscript.

# Supplementary Figures

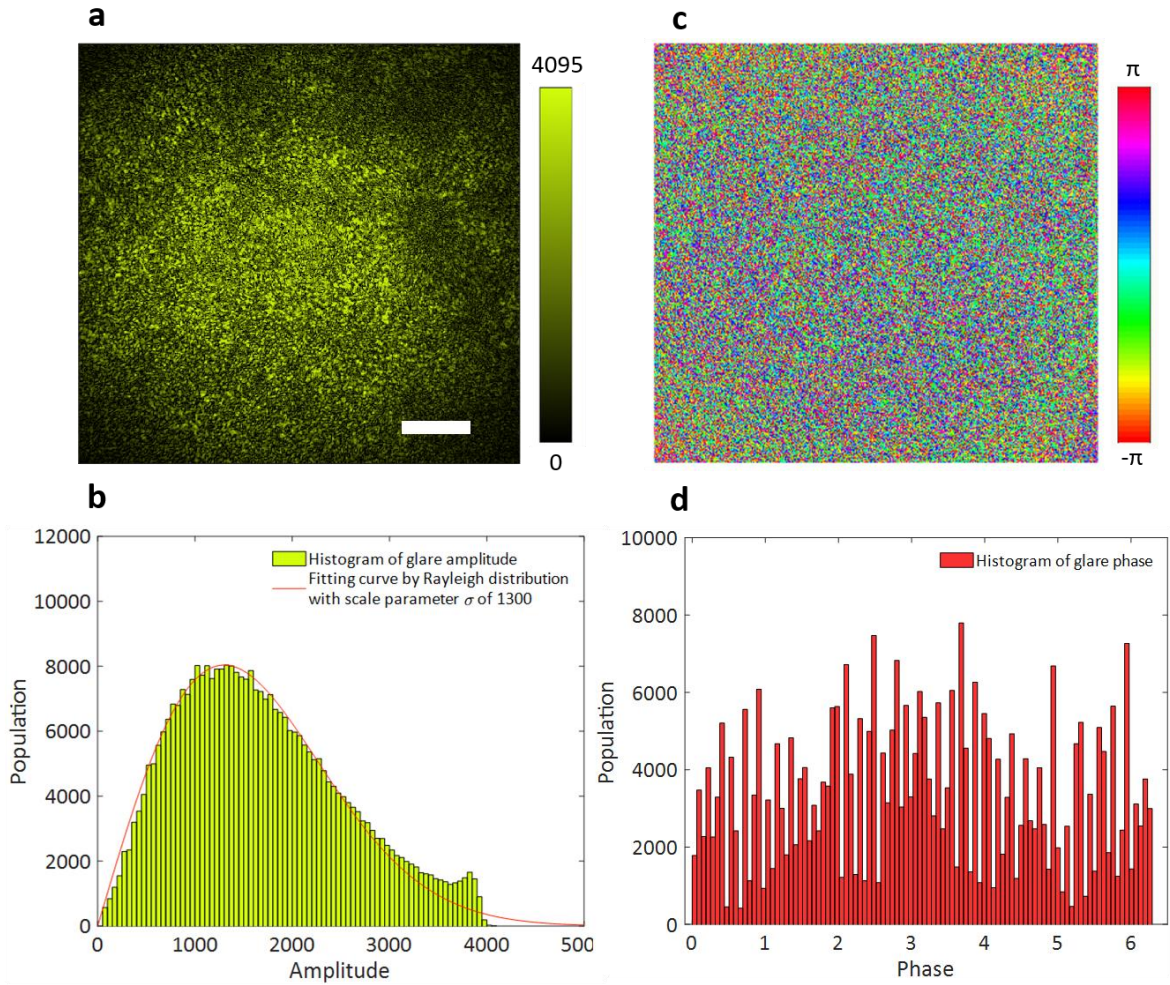

**Supplementary Figure 1 | Glare back-reflected from a scattering medium.** (a) Amplitude distribution of the glare. The amplitude has a speckle distribution caused by the scattering medium. (b) Histogram of the amplitude of the glare. The amplitude of the speckle typically follows a Rayleigh distribution $\frac{A}{\sigma^2}e^{-\frac{A^2}{2\sigma^2}}$ where $A$ is the amplitude and $\sigma$ is a scale parameter. Fitting the data with a Rayleigh distribution ($\sigma$ = 1300) shows good agreement with the histogram of the measured amplitude. (c) Phase distribution of the glare. (d) Histogram of the phase of the glare. The phase is homogeneously distributed over 0 to 2π. Scale bar is 500 μm.

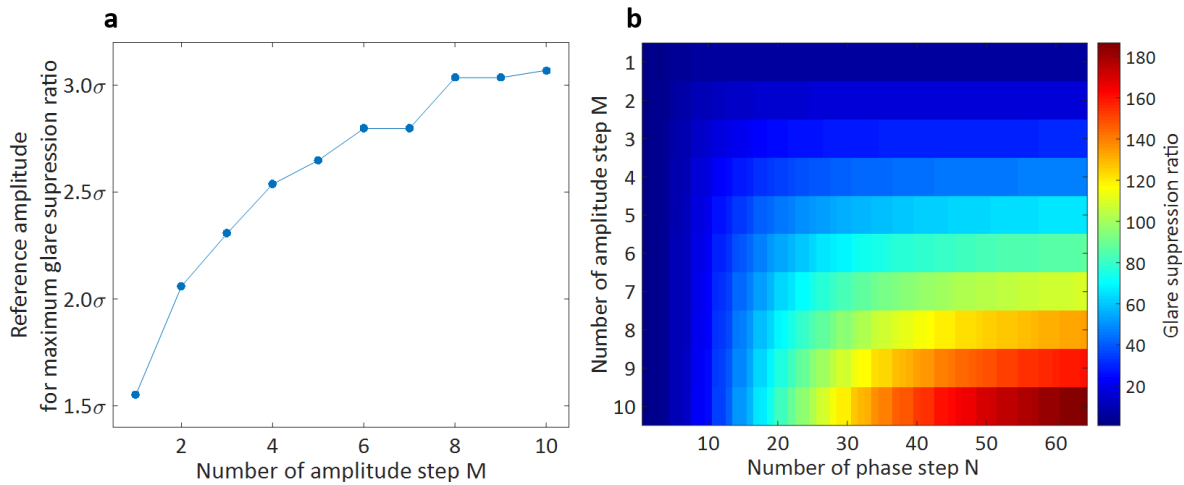

**Supplementary Figure 2 | Ideal glare suppression factor computed via simulation.** To simulate glare, a speckle field of $10^6$ pixels is generated, each of which has an amplitude following a Rayleigh distribution and phase that is uniformly distributed from 0 to $2\pi$. We also generate multiple sets of reference fields each consisting of different numbers of steps in amplitude and phase. For a single set, where the number of steps in amplitude and phase are M and N respectively, the size of all the permutations is M*N. By screening for the minimum value of destructive interference between the speckle field and the whole set of reference fields, the minimum residue of glare in the speckle field is determined as the glare after CGN is applied. The glare suppression factor is calculated by the ratio of the glare intensity before and after CGN is applied. (a) Optimum reference amplitude as a function of the number of amplitude steps. If the reference amplitude when no amplitude modulation is applied is set to the maximum glare amplitude, the glare suppression factor will be extremely low. This is because due to the Rayleigh distributed amplitude, the majority of the glare amplitude values are much lower than its maximum, as shown in Supplementary Fig. 1(b). To efficiently suppress the glare, the reference amplitude must be chosen properly. (b) Glare suppression factor as a function of the number of steps in the reference amplitude and phase. When the number of steps in amplitude and phase are 10 and 32 respectively, the ideal glare suppression factor is around 130.

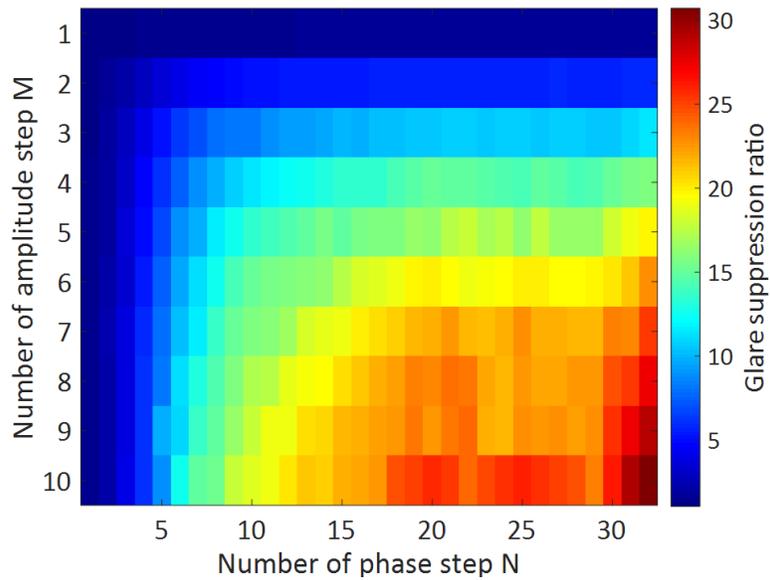

**Supplementary Figure 3 | Measured glare suppression factor.** The mismatch between measured and ideal glare suppression factor can be attributed to: a) phase jitter in the reference beam and the sample beam due to vibration in the system, b) noise in the electronics including the laser and electro-optical modulator, and c) limited extinction ratio of the amplitude modulator, polarized optics, etc. When the number of steps in the amplitude and phase are 10 and 32 respectively, the measured glare suppression factor is around 30.